\documentclass[aps,pra,reprint,showpacs,floatfix,longbibliography]{revtex4-1}
\usepackage{graphicx, amsmath, amssymb}
\usepackage{hyperref}
\usepackage[caption=false]{subfig}
\allowdisplaybreaks[1] 

\begin{document}


\newcommand{\braket}[2]{{\left\langle #1 \middle| #2 \right\rangle}}
\newcommand{\bra}[1]{{\left\langle #1 \right|}}
\newcommand{\ket}[1]{{\left| #1 \right\rangle}}
\newcommand{\ketbra}[2]{{\left| #1 \middle\rangle \middle \langle #2 \right|}}
\newcommand{\fref}[1]{Fig.~\ref{#1}}
\newcommand{\tref}[1]{Table~\ref{#1}}


\title{Isolated Vertices in Continuous-Time Quantum Walks on Dynamic Graphs}

\author{Thomas G.~Wong}
	\email{thomaswong@creighton.edu}
	\affiliation{Department of Physics, Creighton University, 2500 California Plaza, Omaha, NE 68178}

\begin{abstract}
	It was recently shown that continuous-time quantum walks on dynamic graphs, i.e., sequences of static graphs whose edges change at specific times, can implement a universal set of quantum gates. This result treated all isolated vertices as having self-loops, so they all evolved by a phase under the quantum walk. In this paper, we permit isolated vertices to be loopless or looped, and loopless isolated vertices do not evolve at all under the quantum walk. Using this distinction, we construct simpler dynamic graphs that implement the Pauli gates and a set of universal quantum gates consisting of the Hadamard, $T$, and CNOT gates, and these gates are easily extended to multi-qubit systems. For example, the $T$ gate is simplified from a sequence of six graphs to a single graph, and the number of vertices is reduced by a factor of four. We also construct a generalized phase gate, of which $Z$, $S$, and $T$ are specific instances. Finally, we validate our implementations by numerically simulating a quantum circuit consisting of layers of one- and two-qubit gates, similar to those in recent quantum supremacy experiments, using a quantum walk.
\end{abstract}

\pacs{03.67.Ac, 03.67.Lx}

\maketitle


\section{Introduction}

The continuous-time quantum walk is a quantum mechanical analogue of a classical continuous-time random walk. It was introduced in \cite{FG1998a} as a method for solving decision trees, and it has since been applied to a variety of computational problems, such as searching \cite{CG2004} and solving boolean formulas \cite{FGG2008}. Exponential speedups have even been achieved using the continuous-time quantum walk \cite{Childs2003}. Furthermore, they are universal for quantum computing, meaning any quantum computation can be formulated as a continuous-time quantum walk \cite{Childs2009}.

In a quantum walk, the $N$ vertices of a graph represent orthonormal basis states $\ket{0}, \ket{1}, \dots, \ket{N-1}$ of an $N$-dimensional Hilbert space, and the edges of the graph specify the allowed transitions between basis states. In a continuous-time quantum walk, the state of the walker is a quantum superposition over the vertices, and it evolves according to Schr\"odinger's equation with a Hamiltonian either proportional to the Laplacian of the graph \cite{FG1998a}, or proportional to the adjacency matrix $A$ of the graph \cite{Bose2009}. The adjacency matrix is an $N \times N$ matrix that encodes the structure of the graph; $A_{ij} = 1$ if vertices $i$ and $j$ are adjacent, and $A_{ij} = 0$ otherwise. When the graph is regular, the Laplacian and adjacency matrix effect the same evolution (up to a global, unobservable phase), but when the graph is irregular, the walks can differ \cite{Wong19}. Either way, the edges of the graph are encoded in the Hamiltonian.

\begin{figure}
\begin{center}
	\includegraphics{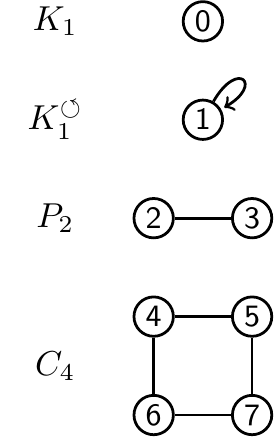}
	\caption{\label{fig:graphs} A loopless isolated vertex ($K_1$), an isolated vertex with a self-loop ($K_1^\circlearrowleft$), a path graph of two vertices ($P_2$), and a cycle of four vertices ($C_4$).}
\end{center}
\end{figure}

In this paper, we focus on the quantum walk effected by the adjacency matrix, such as a single excitation in a spin network with $XY$ interactions \cite{Bose2009}. For convenience and in alignment with several prior works \cite{Kendon2011,HH2019}, we choose the jumping rate to be $-1$, so the Hamiltonian is equal to the adjacency matrix. This corresponds to walking backward in time, but the results can easily be adapted to forward-time evolution. Then for a static graph, the adjacency matrix and Hamiltonian are time-independent, so the solution to Schr\"odinger's equation (with $\hbar = 1$) is
\begin{equation}
	\label{eq:evolution}
	\ket{\psi(t)} = e^{-iAt} \ket{\psi(0)}.
\end{equation}

\begin{table*}
	\caption{\label{table:graphs} Quantum walks on the loopless isolated vertex $K_1$, isolated vertex with a self-loop $K_1^\circlearrowleft$, path graph of two vertices $P_2$, and the cycle of four vertices $C_4$, as depicted in \fref{fig:graphs}. The initial state of the quantum walk is $c_0 \ket{0} + c_1 \ket{1} + \dots + c_7 \ket{7}$.}
\begin{ruledtabular}
\begin{tabular}{cccc}
	Graph & Time & Evolution & Description \\
	\colrule
	$K_1$ & $t$ & $c_0 \ket{0} \to c_0 \ket{0}$ & No evolution \\
	\colrule
	$K_1^\circlearrowleft$ & $t$ & $c_1 \ket{1} \to e^{-it} c_1 \ket{1}$ & Phase \\
	                 & $\pi/2$ & $c_1 \ket{1} \to -i c_1 \ket{1}$ & Phase ($-i$) \\
	                 & $\pi$ & $c_1 \ket{1} \to -c_1 \ket{1}$ & Phase ($-1$) \\
	                 & $3\pi/2$ & $c_1 \ket{1} \to i c_1 \ket{1}$ & Phase ($i$) \\
	                 & $2\pi$ & $c_1 \ket{1} \to c_1 \ket{1}$ & No evolution \\
	\colrule
	$P_2$ & $t$ & $c_2 \ket{2} + c_3 \ket{3} \to [c_2 \cos(t) - i c_3 \sin(t)] \ket{2} + [c_3 \cos(t) - i c_2 \sin(t)] \ket{3}$ & Mix \\
	      & $\pi/4$ & $c_2 \ket{2} + c_3 \ket{3} \to (1/\sqrt{2}) [(c_2 - i c_3) \ket{2} + (c_3 - ic_2) \ket{3}]$ & Mix \\
	      & $\pi/2$ & $c_2 \ket{2} + c_3 \ket{3} \to -i(c_3 \ket{2} + c_2 \ket{3})$ & Swap and phase ($-i$) \\
	      & $\pi$ & $c_2 \ket{2} + c_3 \ket{3} \to -(c_2 \ket{2} + c_3 \ket{3})$ & Phase ($-1$) \\
	      & $3\pi/2$ & $c_2 \ket{2} + c_3 \ket{3} \to i(c_3 \ket{2} + c_2 \ket{3})$ & Swap and phase ($i$) \\
	      & $2\pi$ & $c_2 \ket{2} + c_3 \ket{3} \to c_2 \ket{2} + c_3 \ket{3}$ & No evolution \\
	\colrule
	$C_4$ & $t$ & $\begin{array}{l}
		c_4 \ket{4} + c_5 \ket{5} + c_6 \ket{6} + c_7 \ket{7} \\
		\quad\to (1/2)[c_4-c_7 + (c_4+c_7) \cos(2t) - i (c_5+c_6) \sin(2t)] \ket{4} \\
		\quad\enspace + (1/2)[c_5-c_6 + (c_5+c_6) \cos(2t) - i (c_4+c_7) \sin(2t)] \ket{5} \\
		\quad\enspace + (1/2)[c_6-c_5 + (c_5+c_6) \cos(2t) - i (c_4+c_7) \sin(2t)] \ket{6} \\
		\quad\enspace + (1/2)[c_7-c_4 + (c_4+c_7) \cos(2t) - i (c_5+c_6) \sin(2t)] \ket{7}
	\end{array}$ & Mix \\
	      & $\pi/2$ & $c_4 \ket{4} + c_5 \ket{5} + c_6 \ket{6} + c_7 \ket{7} \to -(c_7 \ket{4} + c_6 \ket{5} + c_5 \ket{6} + c_4 \ket{7})$ & Opposite corners swap and phase ($-1$) \\
	      & $\pi$ & $c_4 \ket{4} + c_5 \ket{5} + c_6 \ket{6} + c_7 \ket{7} \to c_4 \ket{4} + c_5 \ket{5} + c_6 \ket{6} + c_7 \ket{7}$ & No evolution \\
\end{tabular}
\end{ruledtabular}
\end{table*}

For example, \fref{fig:graphs} depicts a graph of eight vertices consisting of four components: a loopless isolated vertex $K_1$, an isolated vertex with a self-loop $K_1^\circlearrowleft$, a path of two vertices $P_2$, and a cycle of four vertices $C_4$. Its adjacency matrix is
\[ A = \begin{pmatrix}
	0 & 0 & 0 & 0 & 0 & 0 & 0 & 0 \\
	0 & 1 & 0 & 0 & 0 & 0 & 0 & 0 \\
	0 & 0 & 0 & 1 & 0 & 0 & 0 & 0 \\
	0 & 0 & 1 & 0 & 0 & 0 & 0 & 0 \\
	0 & 0 & 0 & 0 & 0 & 1 & 1 & 0 \\
	0 & 0 & 0 & 0 & 1 & 0 & 0 & 1 \\
	0 & 0 & 0 & 0 & 1 & 0 & 0 & 1 \\
	0 & 0 & 0 & 0 & 0 & 1 & 1 & 0 \\
\end{pmatrix}. \]
If the initial state of the walker is
\[ \ket{\psi(0)} = c_0 \ket{0} + c_1 \ket{1} + \dots + c_7 \ket{7}, \]
then using \eqref{eq:evolution}, it evolves to the state
\begin{widetext}
\begin{align*} 
	\ket{\psi(t)} 
		&= c_0 \ket{0} + e^{-it} c_1 \ket{1} + [c_2 \cos(t) - i c_3 \sin(t)] \ket{2} + [c_3 \cos(t) - i c_2 \sin(t)] \ket{3} \\
		&\quad+ (1/2)[c_4-c_7 + (c_4+c_7) \cos(2t) - i (c_5+c_6) \sin(2t)] \ket{4} \\
		&\quad+ (1/2)[c_5-c_6 + (c_5+c_6) \cos(2t) - i (c_4+c_7) \sin(2t)] \ket{5} \\
		&\quad+ (1/2)[c_6-c_5 + (c_5+c_6) \cos(2t) - i (c_4+c_7) \sin(2t)] \ket{6} \\
		&\quad+ (1/2)[c_7-c_4 + (c_4+c_7) \cos(2t) - i (c_5+c_6) \sin(2t)] \ket{7}.
\end{align*}
\end{widetext}
From this, we see that the amplitude at the loopless isolated vertex $K_1$ ($\ket{0}$) stays the same. This is summarized in the first row of \tref{table:graphs}. In contrast, the amplitude at the isolated vertex with a self-loop $K_1^\circlearrowleft$ ($\ket{1}$) evolves by a phase $e^{-it}$, and this is summarized in the second row of \tref{table:graphs} with specific times $t = \pi/2, \pi, 3\pi/2, 2\pi$. The amplitudes in the path graph $P_2$ ($\ket{2}$ and $\ket{3}$) mix \cite{Kendon2011}, at times completely swapping (with a phase), as summarized in the third row of \tref{table:graphs}. Finally, the amplitudes in the cycle graph $C_4$ ($\ket{4}$, $\ket{5}$, $\ket{6}$, and $\ket{7}$) also mix \cite{Kendon2011}, and at time $\pi/2$, the amplitudes at opposite corners have swapped (with a phase), as summarized in the last row of \tref{table:graphs}.

Herrman and Humble \cite{HH2019} recently constructed sequences of graphs, called dynamic graphs, on which continuous-time quantum walks implement the Pauli $X$, $Y$, and $Z$ gates, along with the Hadamard gate $H$, $T$ gate (fourth root of $Z$), and CNOT. The last three, $\{ H, T, \text{CNOT} \}$ are a universal set of quantum gates \cite{NielsenChuang2000}, so they can approximate any unitary to any desired precision. Thus, using dynamic graphs that correspond to various quantum gates, continuous-time quantum walks can implement any quantum computation. We note the quantum approximate approximation algorithm (QAOA) \cite{Farhi2014a} also changes the Hamiltonian at discrete times, corresponding to turning on and off interactions. Quantum walks on dynamic graphs are similar.

Herrman and Humble's formulation treated all isolated vertices as having self-loops (i.e., as $K_1^\circlearrowleft$'s), so their amplitudes evolved by phases. Physically, they evolve ``at an energy offset from the other vertices (like a physical mode at a different frequency or in the presence of a different bias)'' \cite{Humble2019}. Simple graphs, however, have no self-loops. Even the $XY$ spin model that gives rise to a quantum walk governed by the adjacency matrix assumes a simple graph \cite{Bose2009}. This raises the question of how permitting loopless isolated vertices in dynamic graphs affect the construction of quantum gates. Physically, they evolve at zero energy, with no phase.

In this paper, we permit isolated vertices to be loopless ($K_1$'s) or looped ($K_1^\circlearrowleft$'s), allowing us to engineer whether the amplitude at a vertex remains constant or evolves by a phase. With this distinction, we are able to significantly reduce the complexity of the dynamic graphs in many cases. By eliminating ancillas, we reduce the $Y$, $Z$, $H$, and $T$ gates from quantum walks on eight vertices to just two vertices. Furthermore, the $H$ gate is reduced a sequence of five graphs to three graphs, and the $T$ gate is reduced from six graphs to just one graph. We also construct a generalized phase gate, of which $Z$, $S$, and $T$ are specific instances. Our implementations also easily extend to multi-qubit systems.

In the next section, we review Herrman and Humble's dynamic graphs for the Pauli gates, and we propose improvements utilizing loopless isolated vertices. Then in Section III, we review and improve dynamic graphs for the universal gate set $\{ H, T, \text{CNOT} \}$ and propose a dynamic graph that implements an arbitrary-phase gate. Following, in Section IV, we validate our constructions by implementing a three-qubit quantum circuit that alternates between layers of one- and two-qubit gates, similar to those used in quantum supremacy experiments \cite{Aaronson2017,Google2019}, as a quantum walk. Finally, we end with concluding remarks.


\section{Pauli Gates}

In this section, we review Herrman and Humble's constructions of the Pauli $X$, $Y$, and $Z$ gates, where isolated vertices all have self-loops. Along the way, we give simpler constructions when loopless isolated vertices are permitted and show how they generalize to multi-qubit systems.

\begin{table*}
\caption{\label{table:Pauli} The Pauli gates implemented by continuous-time quantum walks on dynamic graphs.}
\begin{ruledtabular}
\begin{tabular}{ccc}
	Gate & Herrman and Humble's Implementation & With Loopless Isolated Vertices \\
	\colrule
	$X$ & \raisebox{-0.5\height}{\includegraphics{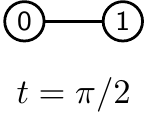} \hspace{0.25in} \includegraphics{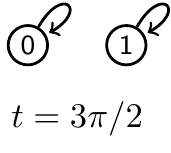}} & \raisebox{-0.5\height}{\includegraphics{X_HH1} \hspace{0.25in} \includegraphics{X_HH2}} \\
		& Total time: $2\pi$ & Total time: $2\pi$ \\
	\colrule
	$Y$ & \raisebox{-0.5\height}{\includegraphics{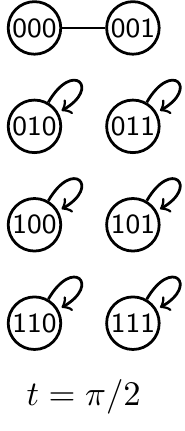} \hspace{0.25in} \includegraphics{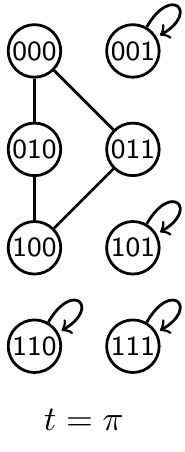}} & \raisebox{-0.5\height}{\includegraphics{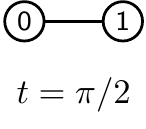} \hspace{0.25in} \includegraphics{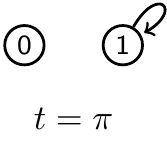}} \\
		& Total time: $3\pi/2$ & Total time: $3\pi/2$ \\
	\colrule
	$Z$ & \raisebox{-0.5\height}{\includegraphics{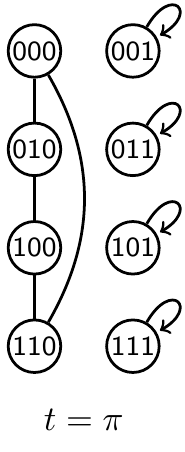}} & \raisebox{-0.5\height}{\includegraphics{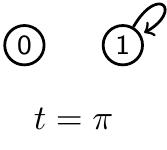}} \\
		& Total time: $\pi$ & Total time: $\pi$ \\
\end{tabular}
\end{ruledtabular}
\end{table*}

Recall the Pauli gates are single-qubit gates. In the computational basis, the $X$ gate acts by $X \ket{0} = \ket{1}$ and $X \ket{1} = \ket{0}$. Then, for a general qubit in a superposition of $\ket{0}$ and $\ket{1}$, 
\[ X ( c_0 \ket{0} + c_1 \ket{1} ) = c_1 \ket{0} + c_0 \ket{1}. \]
To implement this using a quantum walk, Herrman and Humble proposed the dynamic graph shown in the first row of \tref{table:Pauli}. It uses the fact that evolving by $P_2$ for time $\pi/2$ causes the amplitudes at two vertices to swap, but with a phase of $-i$. That is, from the third row of \tref{table:graphs}, $c_0 \ket{0} + c_1 \ket{1} \to -i(c_1 \ket{0} + c_0 \ket{1})$. To remove the phase, each vertex can then evolve in isolation with a self-loop (i.e., $K_1^\circlearrowleft$'s) for time $3\pi/2$, which from the second row of \tref{table:graphs} multiplies their amplitudes by $i$, resulting in $c_1 \ket{0} + c_0 \ket{1}$, hence implementing the $X$ gate. Altogether, these two static graphs ($P_2$ and isolated $K_1^\circlearrowleft$'s) evolve for a combined total time of $2\pi$. This implementation is so direct that loopless isolated vertices do not yield an obvious improvement, so we simply continue using Herrman and Humble's implementation, as shown in the first row of \tref{table:Pauli}.

\begin{figure}
\begin{center}
	\includegraphics{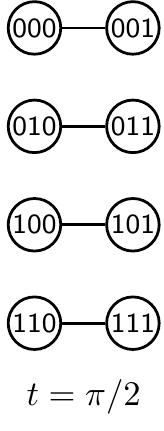} \hspace{0.25in} \includegraphics{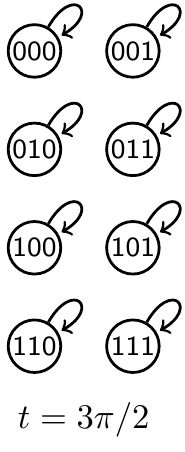}
	\caption{\label{fig:X_multi} A dynamic graph that applies the Pauli $X$ gate to the rightmost of three qubits, i.e., $I \otimes I \otimes X$.}
\end{center}
\end{figure}

This implementation of the $X$ gate is easily extended to multiple qubits. For example, with three qubits, there are eight vertices, which we label with bit strings $\ket{000}, \ket{001}, \dots, \ket{111}$. We can extend Herrman and Humble's dynamic graph so that it applies the $X$ gate to the rightmost qubit by pairing vertices with labelings $\ket{a b 0}$ and $\ket{a b 1}$, where $a$ and $b$ are bits, as shown in \fref{fig:X_multi}. Then we evolve as before with isolated vertices with self-loops. To instead apply $X$ to the leftmost qubit, we would pair vertices with labelings $\ket{0 a b}$ and $\ket{1 a b}$. Finally, to apply $X$ to the middle qubit, we would pair vertices with labelings $\ket{a 0 b}$ and $\ket{a 1 b}$.

Next, the Pauli $Y$ gate acts on the computational basis states by $Y \ket{0} = i\ket{1}$ and $Y \ket{1} = -i\ket{0}$. Then,
\[ Y ( c_0 \ket{0} + c_1 \ket{1} ) = -i c_1 \ket{0} + i c_0 \ket{1}. \]
Herrman and Humble constructed a dynamic graph that implements this, but it requires three ancilla vertices, or five total. Since five vertices requires three qubits, we have drawn all eight vertices corresponding to three qubits in the second row of \tref{table:Pauli}. Say the initial state of the walker is $c_0 \ket{000} + \dots + c_7 \ket{111}$. The first static graph has $\ket{000}$ and $\ket{001}$ evolving by $P_2$ for time $\pi/2$, which from \tref{table:graphs} results in $-i c_1 \ket{000} - i c_0 \ket{001}$ for the first two vertices, while the remaining vertices evolve by a phase of $-i$. Then $\ket{000}$ is kept the same by linking it in $C_4$ to three ancillas for time $\pi$, while $\ket{001}$ and the remaining vertices evolve by a phase of $-1$. The net result is
\begin{align*}
	&{-i} c_1 \ket{000} + i c_0 \ket{001} - i c_2 \ket{010} - i c_3 \ket{011} \\
	&-i c_4 \ket{100} + i c_5 \ket{101} + i c_6 \ket{110} + i c_7 \ket{111}.
\end{align*}
Thus, $\ket{000}$ and $\ket{001}$ underwent a $Y$ gate. The remaining vertices only evolved by phases, but since these ancillas begin with zero amplitude, they also end with zero amplitude, so they do not interfere with larger quantum computations. Note the bit-string labeling of the vertices is confusing in this formulation, since the $Y$ gate is \emph{not} applied to any one of the three qubits of the labeling. Thus, for true multi-qubit systems where we want to apply $Y$ to one of the qubits, a direct generalization of Herrman and Humble's construction requires three ancillas for every $P_2$ pair, although simplifications may be possible by making $C_4$ cycles with other non-ancilla qubits. In contrast, our formulation, next, requires no ancillas.

Permitting loopless isolated vertices, we simplify the $Y$ gate so that only two vertices are required. Herrman and Humble's construction required ancillas so that in the second graph, $\ket{000}$ could be linked in a four cycle $C_4$ that does not evolve in time $\pi$. Now, $\ket{000}$ can be kept from evolving by making it isolated and loopless, eliminating the need for the ancillas. The resulting simpler dynamic graph is shown in the second row of \tref{table:Pauli}. Furthermore, this can be extended to multi-qubit systems in a similar manner to the $X$ gate in \fref{fig:X_multi}. For example, with three qubits, we apply the $Y$ gate to the rightmost qubit by first pairing vertices $\ket{a b 0}$ and $\ket{a b 1}$ in $P_2$'s, followed by making each $\ket{a b 0}$ a $K_1$ and each $\ket{a b 1}$ a $K_1^\circlearrowleft$.

Now, the Pauli $Z$ gate acts on the computational basis by $Z \ket{0} = \ket{0}$ and $Z \ket{1} = -\ket{1}$, so
\[ Z ( c_0 \ket{0} + c_1 \ket{1} ) = c_0 \ket{0} - c_1 \ket{1}. \]
So the goal of the $Z$ gate is to leave $\ket{0}$ unchanged while changing the phase of $\ket{1}$. To keep $\ket{0}$ unchanged, Herrman and Humble again utilized three ancillas, or five vertices total. Again, this requires three qubits, or eight vertices, which we draw in the last row of \tref{table:Pauli}. For time $\pi$, $\ket{000}$ is kept in a four cycle $C_4$ with $\ket{010}$, $\ket{100}$, and $\ket{110}$, while the remaining vertices evolve with a phase of $-1$. That is, $c_0 \ket{000} + \dots + c_7 \ket{111}$ evolves to
\begin{align*}
	&c_0 \ket{000} - c_1 \ket{001} + c_2 \ket{010} - c_3 \ket{011} \\
	&+ c_4 \ket{100} - c_5 \ket{101} + c_6 \ket{110} - c_7 \ket{111}.
\end{align*}
Thus, $\ket{000}$ and $\ket{001}$ underwent a $Z$ gate, with the ancillas retaining their initial amplitudes of zero. If the ancillas have nonzero initial amplitude, however, note this dynamic graph applies the $Z$ gate to the rightmost qubit (i.e., $I \otimes I \otimes Z$), so the labeling has some meaning here. Next, our construction requires no ancillas.

With loopless isolated vertices, we simplify the $Z$ gate by eliminating the need for ancillas and the cycle $C_4$. $\ket{0}$ can be made stationary under $K_1^\circlearrowleft$ while $\ket{1}$ evolves by a phase under $K_1$. This is depicted in the last row of \tref{table:Pauli}. As with our implementations of the $X$ and $Y$ gates, our $Z$ gate naturally extends to multi-qubit systems in a similar manner to the $X$ gate in \fref{fig:X_multi}.

For the identity gate, Herrman and Humble gave several possible constructions, including evolving with all isolated vertices with self-loops $K_1^\circlearrowleft$ for time $2\pi$, pairs of vertices $P_2$ for time $2\pi$, or cycles of four vertices $C_4$ for time $\pi$. Now, we can implement the identity gate using loopless isolated vertices, which do not evolve at all, no matter the time. This makes it easier to see which vertices are actively evolving and which are not.


\section{Universal Quantum Gates}

\begin{table*}
\caption{\label{table:universal} A set of universal quantum gates implemented by continuous-time quantum walks on dynamic graphs.}
\begin{ruledtabular}
\begin{tabular}{ccc}
	Gate & Herrman and Humble's Implementation & With Loopless Isolated Vertices \\
	\colrule
	$H$ & \begin{minipage}{3in}
		\includegraphics{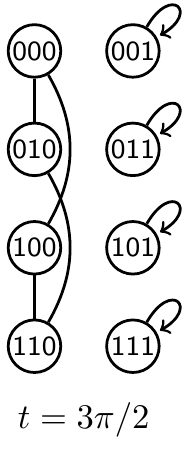} \hspace{0.25in} \includegraphics{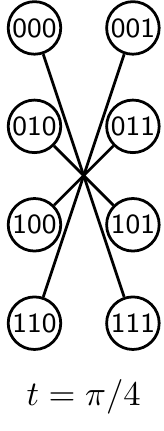} \hspace{0.25in} \includegraphics{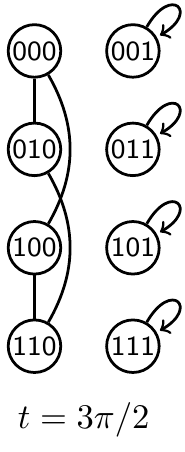} \\
		\includegraphics{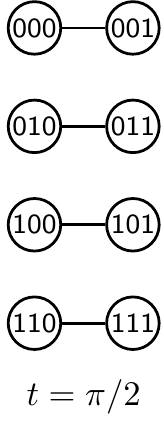} \hspace{0.25in} \includegraphics{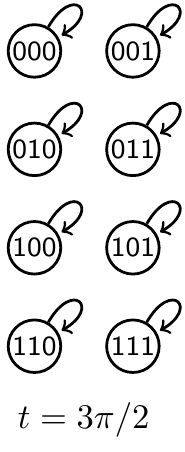}
	\end{minipage} & \raisebox{-0.5\height}{\includegraphics{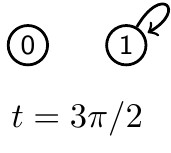} \hspace{0.25in} \includegraphics{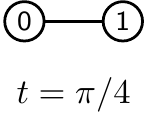} \hspace{0.25in} \includegraphics{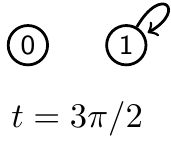}} \\
		& Total time: $21\pi/4$ & Total time: $13\pi/4$ \\
	\colrule
	$T$ & \begin{minipage}{3in}
		\includegraphics{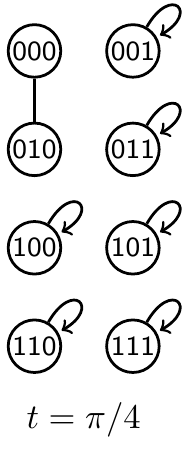} \hspace{0.25in} \includegraphics{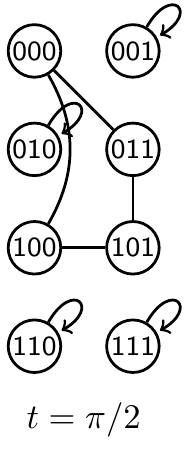} \hspace{0.25in} \includegraphics{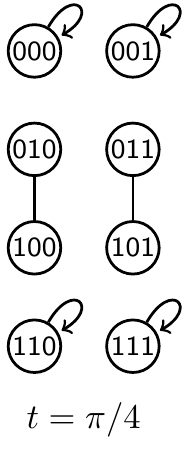} \\
		\includegraphics{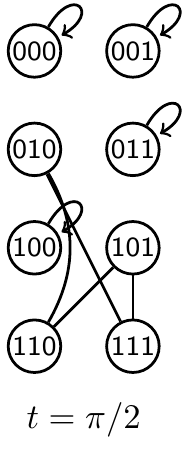} \hspace{0.25in} \includegraphics{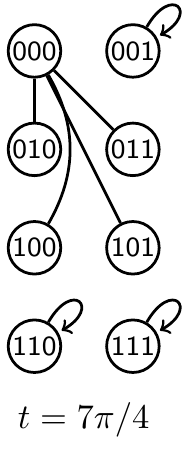} \hspace{0.25in} \includegraphics{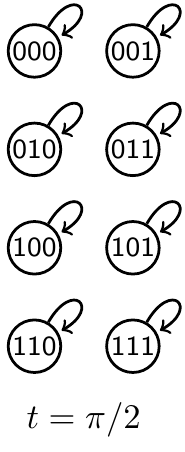}
	\end{minipage} & \raisebox{-0.5\height}{\includegraphics{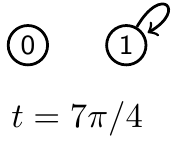}} \\
		& Total time: $15\pi/4$ & Total time: $7\pi/4$\\
	\colrule
	CNOT & \raisebox{-0.5\height}{\includegraphics{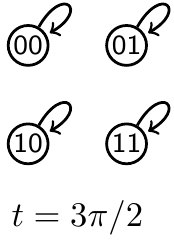} \hspace{0.25in} \includegraphics{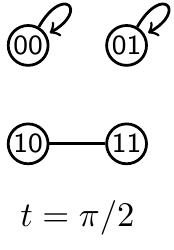}} & \raisebox{-0.5\height}{\includegraphics{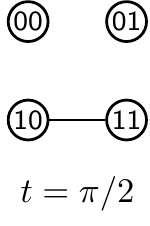} \hspace{0.25in} \includegraphics{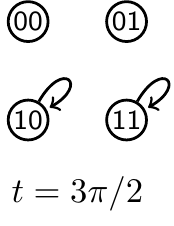}} \\
		& Total time: $2\pi$ & Total time: $2\pi$ \\
\end{tabular}
\end{ruledtabular}
\end{table*}

In this section, we review Herrman and Humble's constructions of the Hadamard $H$, $T$, and CNOT gates, and we propose simpler implementations with loopless isolated vertices. Together, these are a universal set of quantum gates \cite{NielsenChuang2000}.

First, the Hadamard gate is a single-qubit gate, and it acts on computational basis states by $H \ket{0} = (\ket{0} + \ket{1})/\sqrt{2}$ and $H \ket{1} = (\ket{0} - \ket{1})/\sqrt{2}$. Thus, it transforms the superposition $c_0 \ket{0} + c_1 \ket{1}$ to
\begin{equation}
	\label{eq:Hadamard}
	\frac{1}{\sqrt{2}} \left( c_0 + c_1 \right) \ket{0} + \frac{1}{\sqrt{2}} \left( c_0 - c_1 \right) \ket{1}.
\end{equation}
Herrman and Humble's implementation of the Hadamard gate is shown in the first row of \tref{table:universal}. Their implementation uses ancillas for a total of eight vertices or three qubits. As proved in their appendix, if the initial state is $c_0 \ket{000} + \dots + c_7 \ket{111}$, their sequence of five graphs transforms this to
\begin{align*}
	&\frac{1}{\sqrt{2}} \left( c_0 + c_1 \right) \ket{000} + \frac{1}{\sqrt{2}} \left( c_0 - c_1 \right) \ket{001} \\
	&+ \frac{1}{\sqrt{2}} \left( c_2 + c_3 \right) \ket{010} + \frac{1}{\sqrt{2}} \left( c_2 - c_3 \right) \ket{011} \\
	&+ \frac{1}{\sqrt{2}} \left( c_4 + c_5 \right) \ket{100} + \frac{1}{\sqrt{2}} \left( c_4 - c_5 \right) \ket{101} \\
	&+ \frac{1}{\sqrt{2}} \left( c_6 + c_7 \right) \ket{110} + \frac{1}{\sqrt{2}} \left( c_6 - c_7 \right) \ket{111}.
\end{align*}
Thus, the Hadamard gate has been applied to $\ket{000}$ and $\ket{001}$, with the six ancillas starting and ending with zero amplitude. Alternatively, if the ancillas have nonzero amplitude, this applies the Hadamard gate to the rightmost of the three qubits, and the bit-string labeling has a natural interpretation. Next, our construction requires no ancillas.

Permitting loopless isolated vertices, we can implement the Hadamard gate using just two vertices, or one qubit. From the third row of \tref{table:graphs}, recall evolution by $P_2$ for time $\pi/4$ transforms $c_0 \ket{0} + c_1 \ket{1}$ to
\[ \frac{1}{\sqrt{2}} (c_0 - i c_1) \ket{0} + \frac{1}{\sqrt{2}} (-ic_0 + c_1) \ket{1}. \]
This is very similar to the Hadamard transform \eqref{eq:Hadamard}, except for some imaginary components. To get the correct phases, we start with $c_0 \ket{0} + c_1 \ket{1}$ and put a self-loop on $\ket{1}$. Evolving for time $3\pi/2$, this yields
\[ c_0 \ket{0} + i c_1 \ket{1}. \]
Next, we connect the two vertices in a path and evolve for time $\pi/4$, which yields
\[ \frac{1}{\sqrt{2}} \left( c_0 + c_1 \right) \ket{0} - \frac{i}{\sqrt{2}} \left( c_0 - c_1 \right) \ket{1}. \]
Finally, to eliminate the factor of $-i$ in the second term, we evolve $\ket{1}$ with a self-loop for time $3\pi/2$:
\[ \frac{1}{\sqrt{2}} \left( c_0 + c_1 \right) \ket{0} + \frac{1}{\sqrt{2}} \left( c_0 - c_1 \right) \ket{1}. \]
Thus, we have applied the Hadamard gate. This prescription is depicted in the first row of \tref{table:universal}. It reduced the Hadamard transform from a sequence of five graphs of eight vertices each to a sequence of three graphs of two vertices each. It is also easily extended to multiple qubits in a similar fashion to the $X$ gate in \fref{fig:X_multi}.

\begin{figure}
\begin{center}
	\includegraphics{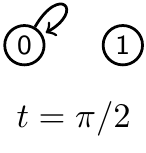} \hspace{0.15in} \includegraphics{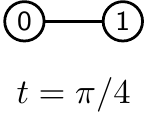} \hspace{0.15in} \includegraphics{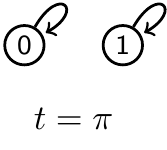} \hspace{0.15in} \includegraphics{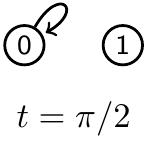}

	Total time: $9\pi/4$
	\caption{\label{fig:H_alt} An alternate implementation of the Hadamard gate with loopless isolated vertices.}
\end{center}
\end{figure}

Our construction of $H$ also reduced the total time from $21\pi/4$ to $13\pi/4$. It can be further reduced to $9\pi/4$, as shown in \fref{fig:H_alt}, but it has a fourth graph in the sequence. First, for time $\pi/2$, $\ket{0}$ evolves as $K_1^\circlearrowleft$ while $\ket{1}$ remains constant as $K_1$, which evolves $c_0 \ket{0} + c_1 \ket{1}$ to $-i c_0 \ket{0} + c_1 \ket{1}$. Then they evolve as $P_2$ for time $\pi/4$, resulting in $-i(c_0 + c_1)/\sqrt{2} \ket{0} - (c_0 - c_1)/\sqrt{2} \ket{1}$. Third, both $\ket{0}$ and $\ket{1}$ evolve as $K_1^\circlearrowleft$'s for time $\pi$, negating the state to $i(c_0 + c_1)/\sqrt{2} \ket{0} + (c_0 - c_1)/\sqrt{2} \ket{1}$. Finally, $\ket{0}$ evolves as $K_1^\circlearrowleft$ while $\ket{1}$ remains constant as $K_1$ for time $\pi/2$, finishing the Hadamard transform \eqref{eq:Hadamard}. Altogether, these four graphs take time $9\pi/4$.

These times assume that the Hamiltonian is equal to the adjacency matrix. Then larger, more connected graphs use more energy since the Hamiltonian has a larger norm. Reducing the energy corresponds to increasing the runtime, so with constant energy, our speedup is even greater because we reduced the complexity of the graphs. This consideration also holds for the $Y$ and $Z$ gates previously discussed.

Next, the $T$ gate is also a single-qubit gate, and it acts on computational basis states by $T \ket{0} = \ket{0}$ and $T \ket{1} = e^{i\pi/4} \ket{1}$. Thus,
\[ T ( c_0 \ket{0} + c_1 \ket{1}) = c_0 \ket{0} + e^{i\pi/4} c_1 \ket{1}. \]
Herrman and Humble implemented this with the help of six additional vertices, so the eight total vertices correspond to three qubits. Their dynamic graph, which consists of six static graphs, is shown in the second row of \tref{table:universal}. Besides the graphs summarized in \tref{table:graphs}, it also includes the star graph of five vertices in the fifth graph. As proved in their appendix, it evolves $c_0 \ket{000} + \dots + c_7 \ket{111}$ to
\begin{align*}
	&c_0 \ket{000} + e^{i\pi/4} c_1 \ket{001} \\
	&\quad- \frac{1}{2} \left( c_2 - \sqrt{2} e^{-i\pi/4} c_4 + c_5 \right) \ket{010} \\
	&\quad- \frac{1}{2} \left( c_2 + \sqrt{2} e^{-i\pi/4} c_4 + c_5 \right) \ket{011} \\
	&\quad+ \frac{1}{2} \left( c_2 + \sqrt{2} e^{-i\pi/4} c_3 + c_5 \right) \ket{100} \\
	&\quad+ \frac{1}{2} \left( c_2 - \sqrt{2} e^{-i\pi/4} c_3 + c_5 \right) \ket{101} \\
	&\quad+ e^{-i\pi/4} c_7 \ket{110} + e^{-i\pi/4} c_6 \ket{111}.
\end{align*}
Thus, the $T$ gate was applied to $\ket{000}$ and $\ket{001}$, with the ancillas retaining their initial zero amplitudes. As with Herrman and Humble's $Y$ gate, the labeling is a little confusing because this does not correspond to the $T$ gate acting on any of three three qubits of the labeling.

To extend Herrman and Humble's result to multi-qubit systems, we can add six ancilla vertices to each pair of vertices. For example, say we have two qubits with vertices $\ket{00}$, $\ket{01}$, $\ket{10}$, and $\ket{11}$. To apply $T$ to the qubit on the right, we construct Herrman and Humble's eight-vertex dynamic graph using $\ket{00}$, $\ket{01}$, and six ancilla vertices, and in parallel, we do the same with $\ket{10}$, $\ket{11}$, and another six ancilla vertices, for a total of twelve ancillas. The quantum walk on this multiplies $\ket{01}$ and $\ket{11}$ by $e^{i\pi/4}$ while leaving $\ket{00}$ and $\ket{10}$ unchanged, hence applying the $T$ gate to the right qubit. The ancillas keep their initially zero amplitude. With $n$ qubits, this generalization requires $2^{n+2}$ vertices to apply the $T$ gate to a single qubit. We can also modify this algorithm so that each pair (e.g., $\ket{00}$ and $\ket{01}$, or $\ket{10}$ and $\ket{11}$) gets $T$ applied sequentially. Then, the six ancillas can be reused, but the computational time is multiplied by $2^{n-1}$.

With loopless isolated vertices, the $T$ gate becomes trivial. $\ket{0}$ can be made static using a loopless isolated vertex, and $\ket{1}$ can be given a phase of $e^{i\pi/4}$ using a self-loop for time $7\pi/4$. This is shown in the second row of \tref{table:universal}, and it is a dramatic reduction from six graphs of eight vertices each to a single graph of two vertices. The total time is also more than halved from $15\pi/4$ to $7\pi/4$. Finally, our simpler construction is easily extended to multiple qubits, similar to \fref{fig:X_multi}. Compared to the parallel generalization of Herrman and Humble's approach, our construction reduces the number of vertices by a factor of four (from eight vertices for each pair to simply the pair).

This construction of $T$ can also be modified to implement any phase $e^{i\theta}$ by evolving for time $n(2\pi) - \theta$ for integer $n$ such that the evolution time is positive. For example, to implement the phase gate $S$ (the square root of $Z$), we can evolve for time $1 \cdot 2\pi - \pi/2 = 3\pi/2$.

Closing out the universal gate set, CNOT is a two-qubit gate that flips the second qubit if the first qubit is 1. Then, it transforms $c_0 \ket{00} + \dots + c_3 \ket{11}$ to
\[ c_0 \ket{00} + c_1 \ket{01} + c_3 \ket{10} + c_2 \ket{11}. \]
In other words, we simply swap the amplitudes at $\ket{10}$ and $\ket{11}$, which can be done using $P_2$ for time $\pi/2$, but this includes an overall phase of $-i$. To remove the phase, Herrman and Humble begin by evolving with isolated vertices with self-loop for time $3\pi/2$, as shown in the last row of \tref{table:universal}. Note vertices $\ket{00}$ and $\ket{01}$ evolve the entire time of $2\pi$ as isolated vertices with self-loops, which means there is no net change. It may be more clear to make them loopless isolated vertices, however, so that regardless the total evolution time, their amplitudes remain constant. This is shown in the last row of \tref{table:universal}. Note we also swapped the order of the graphs for clarity. Finally, as noted by Herrman and Humble, the Toffoli gate has a similar implementation, except it swaps $\ket{110}$ and $\ket{111}$.


\section{Circuit Simulation}

\begin{figure}
\begin{center}
	\includegraphics{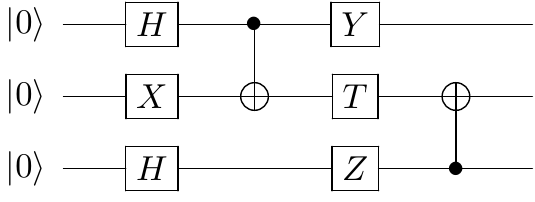}
	\caption{\label{fig:layers} A quantum circuit of three qubits with four alternating layers of one- and two-qubit gates.}
\end{center}
\end{figure}

\begin{figure*}
\begin{center}
	\includegraphics{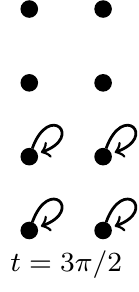} \hspace{0.25in}
	\includegraphics{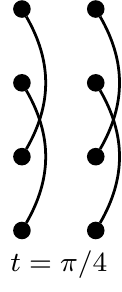} \hspace{0.25in}
	\includegraphics{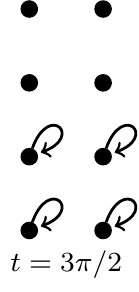} \hspace{0.25in}
	\includegraphics{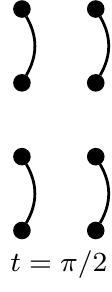} \hspace{0.25in}
	\includegraphics{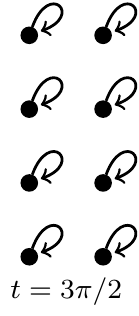} \hspace{0.25in}
	\includegraphics{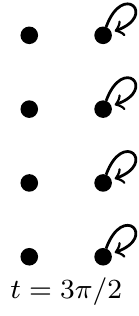} \hspace{0.25in}
	\includegraphics{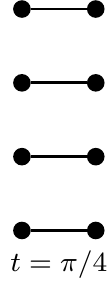} \hspace{0.25in}
	\includegraphics{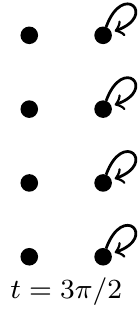}

	\includegraphics{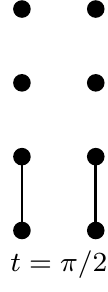} \hspace{0.25in}
	\includegraphics{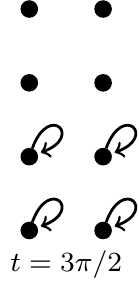} \hspace{0.25in}
	\includegraphics{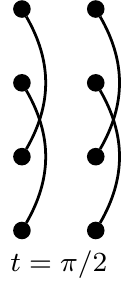} \hspace{0.25in}
	\includegraphics{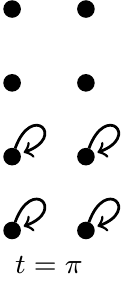} \hspace{0.25in}
	\includegraphics{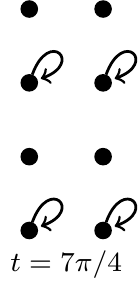} \hspace{0.25in}
	\includegraphics{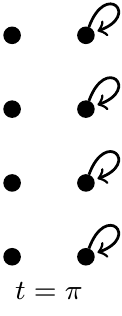} \hspace{0.25in}
	\includegraphics{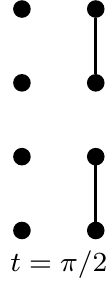} \hspace{0.25in}
	\includegraphics{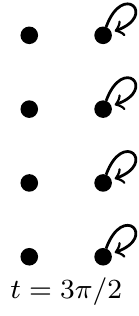}

	\caption{\label{fig:layers_dynamic} A dynamic graph on which a continuous-time quantum walk implements the circuit in \fref{fig:layers}. The eight vertices $000, 001, \dots, 111$ are ordered from top-to-bottom, left-to-right.}
\end{center}
\end{figure*}

\begin{figure}
\begin{center}
	\subfloat[] {
		\label{fig:layers_walk1}
		\includegraphics{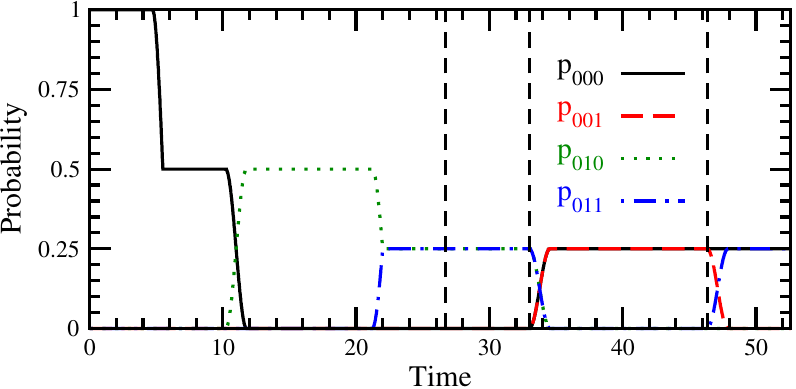}
	}

	\subfloat[] {
		\label{fig:layers_walk2}
		\includegraphics{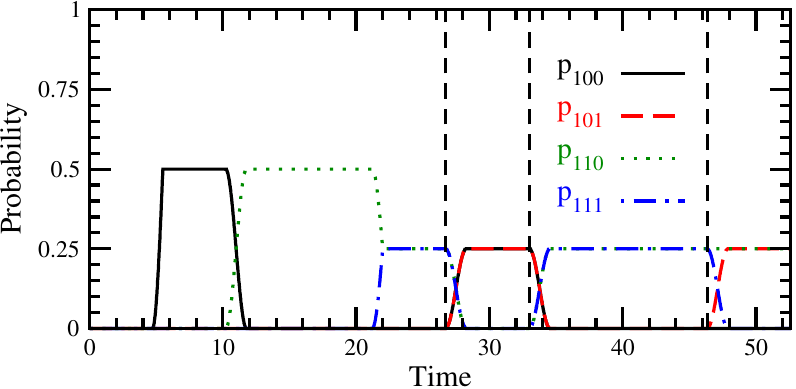}
	}

	\caption{\label{fig:layers_walk} The probability at each vertex as the quantum walk on \fref{fig:layers_dynamic} evolves with time. The vertical dashed lines mark $t = 17\pi/2$, $21\pi/2$, and $59\pi/4$, and the figure ends at $t = 67\pi/4$. In (a), the solid black, dashed red, dotted green, and dot-dashed blue curves denote the probability in vertices $\ket{000}$, $\ket{001}$, $\ket{010}$, and $\ket{011}$, respectively. In (b), they respectively denote the probability in vertices $\ket{100}$, $\ket{101}$, $\ket{110}$, and $\ket{111}$.}
\end{center}
\end{figure}

Now let us verify these implementations by simulating the quantum circuit shown in \fref{fig:layers}. This circuit has four layers that alternate between one-qubit gates and two-qubit gates. Similar random circuits, with more qubits and layers, are used to demonstrate quantum supremacy \cite{Aaronson2017,Google2019}. The qubits begin in the all-zeros state $\ket{000}$, and the state after each layer can be directly calculated:
\begin{align}
	\ket{000} 
		\xrightarrow{H \otimes X \otimes H} & \frac{1}{2} \left( \ket{010} + \ket{011} + \ket{110} + \ket{111} \right) \label{eq:layer1} \\
		\xrightarrow{\text{CNOT}_{1,2}} & \frac{1}{2} \left( \ket{010} + \ket{011} + \ket{100} + \ket{101} \right) \label{eq:layer2} \\
		\xrightarrow{Y \otimes T \otimes Z} & \frac{1}{2} \Big( -i \ket{000} + i \ket{001} \label{eq:layer3} \\
		&\quad + e^{3\pi i/4} \ket{110} - e^{3\pi i/4}\ket{111} \Big) \nonumber \\
		\xrightarrow{\text{CNOT}_{3,2}} & \frac{1}{2} \Big( -i \ket{000} + i \ket{011} \label{eq:layer4} \\
		&\quad - e^{3\pi i/4} \ket{101} + e^{3\pi i/4}\ket{110} \Big). \nonumber
\end{align}
Note the circuit includes all the Pauli $X$, $Y$, $Z$ gates and the entire universal gate set $H$, $T$, CNOT. This differs from Herrman and Humble's simulations of quantum teleportation and the quantum ripple-carry adder, which did not use the $Y$ or $T$ gates. Their simulation of Fig.~\ref{fig:layers} would require ancillas, but ours does not require any; only $2^3 = 8$ vertices are needed.

Using our constructions, the circuit in \fref{fig:layers} is implemented by a quantum walk on the dynamic graph shown in \fref{fig:layers_dynamic}. The first three graphs implement $H$ on the first (leftmost) qubit, graphs four and five implement $X$ on the second qubit, graphs six through eight implement $H$ on the third qubit, graphs nine and ten implement CNOT with the first qubit as the control and the second qubit as the target, graphs eleven and twelve implement $Y$ on the first qubit, graph thirteen implements $T$ on the second qubit, graph fourteen implements $Z$ on the third qubit, and graphs fifteen and sixteen implement CNOT with the third qubit as the control and the second qubit as the target.

Performing a continuous-time quantum walk on this dynamic graph, we get the evolution in \fref{fig:layers_walk}. At $t = 0$, the probability is entirely at vertex $\ket{000}$, as shown by the solid black curve in \fref{fig:layers_walk1}. At $t = 17\pi/2 \approx 26.7$, which is marked by the first vertical dashed line, the first layer of the circuit has been applied, and the states $\ket{010}$, $\ket{011}$, $\ket{110}$, and $\ket{111}$ each have probability $1/4$, as expected from \eqref{eq:layer1}. These correspond to the dotted green curve in \fref{fig:layers_walk1}, the dot-dashed blue curve in \fref{fig:layers_walk1}, the dotted green curve in \fref{fig:layers_walk2}, and the dot-dashed blue curve in \fref{fig:layers_walk2}, respectively. Then, at $t = 21\pi/2 \approx 33.0$, which is marked by the second vertical dashed line, the second layer of the circuit has been applied. In agreement with \eqref{eq:layer2}, the probability is $1/4$ each at vertices $\ket{010}$, $\ket{011}$, $\ket{100}$, and $\ket{101}$, which correspond to the dotted green curve in \fref{fig:layers_walk1}, the dot-dashed blue curve in \fref{fig:layers_walk1}, the solid black curve in \fref{fig:layers_walk2}, and the dashed red curve in \fref{fig:layers_walk2}, respectively. Next, at $t = 59\pi/4 \approx 46.3$, which is marked by the third vertical dashed line, the third layer of the circuit has been applied, and the probability is now $1/4$ at each of the vertices $\ket{000}$, $\ket{001}$, $\ket{110}$, and $\ket{111}$, in agreement with \eqref{eq:layer3}. These probabilities correspond to the solid black curve in \fref{fig:layers_walk1}, dashed red curve in \fref{fig:layers_walk1}, dotted green curve in \fref{fig:layers_walk2}, and dot-dashed blue curve in \fref{fig:layers_walk2}. Finally, at the end of the figures at $t = 67\pi/4 \approx 52.6$, the entire circuit has been applied. In agreement with \eqref{eq:layer4}, the probability is $1/4$ at each of the vertices $\ket{000}$, $\ket{011}$, $\ket{101}$, and $\ket{111}$, which correspond to the solid black curve in \fref{fig:layers_walk1}, dot-dashed blue curve in \fref{fig:layers_walk1}, dashed red curve in \fref{fig:layers_walk2}, and dotted green curve in \fref{fig:layers_walk2}. Furthermore, although they are not shown in the figures, the amplitudes of the simulation also agree with the analytical calculation, with the correct phases.


\section{Conclusion}

To summarize, quantum walks are the basis for many quantum algorithms, just as classical random walks, or Markov chains, are the basis for many classical algorithms. Since quantum walks are universal, they fully encompass the power of quantum computing. Recent work by Herrman and Humble gave a new proof of universality by giving explicit constructions of dynamic graphs that implement a set of universal quantum gates. In this paper, we simplified these constructions by permitting isolated vertices to be loopless, which keeps their amplitudes constant for all time. This eliminated ancillas, reduced the number of graphs in sequences, yielded constructions that are easily generalized to multi-qubit systems, and provided clarity as to which vertices are unchanged in a given step. As such, we have simplified the conversion of quantum circuits to continuous-time quantum walks. We also constructed an arbitrary phase gate, of which $Z$, $S$, and $T$ are specific instances. Finally, we verified our constructions by simulating a quantum circuit that alternates between layers of one-qubit and two-qubit gates, similar to those used to demonstrate quantum supremacy, using a quantum walk. This circuit validated all the Pauli gates and our entire universal gate set.

Further research includes improving the constructions we discussed or constructing dynamic graphs for other quantum gates. One example is to permit self-loops on non-isolated vertices. As a trivial example, if both vertices in $P_2$ also have self-loops, then besides the usual mixing of the amplitudes, each will also be multiplied by a phase $e^{-it}$. This would reduce the time of some gates by combining consecutive $P_2$'s and $K_1^\circlearrowleft$'s for some time, such as in our constructions of the $X$, alternate $H$ in \fref{fig:H_alt}, and CNOT gates. Another approach is using lackadaisical quantum walks \cite{Wong10}, where vertices have self-loops of various weights to indicate how ``lazy'' the walk is at each vertex \cite{Wong27}.


\begin{acknowledgments}
	This work was supported by startup funds from Creighton University.
\end{acknowledgments}


\bibliography{refs}

\end{document}